\documentclass[%
 reprint,
 amsmath,amssymb,
 aps,
]{revtex4-1}
\usepackage{graphicx}
\usepackage{dcolumn}
\usepackage{bm}
\usepackage{hyperref}

\begin{document}

\title{Higher Dimensional Limit Cycles and Coupling Induced Synchronization in Dynamical Systems}

\author{Satadal Datta$^{1,2}$}
\affiliation{$^1$ Harish-Chandra Research Institute, HBNI,\\
  Chhatnag Road, Jhunsi, Allahabad 211 019, India\\
  $^2$  Department of Physics and Astronomy, Seoul National University, 08826 Seoul, Korea}
\author{Jayanta Kumar Bhattacharjee}%
\affiliation{%
  Department of Theoretical Physics, Indian Association for the Cultivation of Science,\\
  Jadavpur, Kolkata 700032,India}%

\author{Dibya Kanti Mukherjee}
\affiliation{
 Laboratoire de Physique des Solides, CNRS UMR 8502,
Universit\'e Paris-Saclay, 91405 Orsay Cedex, France
}%

\begin{abstract}
Limit cycles (attractors for neighbouring periodic orbits in a dissipative dynamical system) have been widely studied but the corresponding generalization for quasi periodic orbits have rarely been discussed. Here we investigate ``higher dimensional limit cycles" by analysing a pair of coupled non identical Van der Pol oscillators and also the Kuznetsov oscillator. We find that the renormalization group based approach introduced by Chen, Goldenfeld and Oono is ideally suited for analysing the quasi periodic analogues of limit cycles. We also address entrainment issues in a pair of forced and coupled Van der Pol oscillators. Our principle finding there is that if two such independent oscillators one with frequency entrainment and the other without are coupled linearly, then it is possible to produce an entrained state via the coupling.
\end{abstract}

\maketitle

\section{Introduction}

A limit cycle, in the study of dynamical systems, is a closed trajectory in phase space having the property that at least one other trajectory spirals into it as time approaches positive or negative infinity. In the case when all neighbouring trajectories approach the limit cycle as time approaches infinity , it is called a stable or attractive limit cycle. Limit cycles appear primarily in two dimensional nonlinear dynamical systems \cite{JS,Strogatz,Drazin_1992}. They have also been located in higher dimensional systems \cite{Lorenz_1963,Rossler_1976,Shahruz_2001,Fortin_2012,Das_2020}. In the study of three or higher dimensional system , the natural extension of the limit cycle would be to a fixed geometrical object on which a trajectory winds around and which has the property that all neighbouring trajectories eventually wind round this object. In four or higher dimensions this object could be a ``fixed torus" which would be a $1\times 1$ circle , characterized by two angles in two mutually perpendicular planes . If the radii of the individual circles are constant in time then we have a ``fixed torus". It should be made clear that these fixed objects have nothing to do with the invariant tori of conservative dynamical systems. The invariant torus that appears there describes the object around which the trajectory winds for a given initial condition and it changes if the initial conditions change. In an interesting recent note, Sprott \cite{Sprott_2014} devised a three dimensional dynamical system where strange attractors and invariant tori co-exist leading to the observation that this dissipative system has a consevative system like behaviour. Interestingly enough in the study of higher dimensional dynamical systems the focus so far has been on locating a limit cycle, rather than to look for ``fixed structures" of higher dimension. Our aim in this work is primarily to look for ``fixed geometrical objects" in higher dimensional dynamical systems by employing the renormalization group technique\cite{Chen_1996,Paquette_2000,Kovalev_1999} of Chen et al\cite{Chen_1996} as a tool.
The usual perturbative paths for studying periodic orbits in dynamical systems has been the Lindstedt -Poincare technique \cite{Landau_1976}, the harmonic balance \cite{Nayfeh_2007}, Krylov-Bogoliubov amplitude equations \cite{Krylov_1950} and the multiple time scale expansion\cite{Nayfeh_2007}. It is obvious that for a quasi-periodic trajectory with two incommensurate frequencies $\omega$ and $\omega'$, the harmonic balance and the Lindstedt -Poincare technique will not work since the frequencies $\omega+\omega'$ and $\omega-\omega'$ will have to be treated at the same footing as the harmonics of the individual frequencies and this cannot be done. In both the Krylov-Bogoliubov method and the multiple time scale expansion,one talks about a 'slow' and a 'fast' time scale in a perturbative analysis with the fast time scale of $\mathcal{O}(\omega,\omega')$ and a slow time scale longer than the fast scale by some inverse power of the small expansion parameter. At some stage these techniques require an averaging over the fast time scale and it would not be clear whether the time scale corresponding to the frequency $n\omega-m\omega'$ ( $m,n$ are integers) is fast or slow. At the heart of the difficulty lies the fact that limit cycles refer to periodic orbits and at some point all the techniques that we mentioned exploit the fact that any periodic function can be expanded in a complete set of linearly independent periodic functions. The lack of such an expansion for a quasi-periodic function is the root cause of difficulties when the traditional techniques are used to study quasi-periodic trajectories. The RG works with a very different philosophy where complete sets do not play a role and is ideally suited for this problem.

We study the coupled non identical Van der Pol oscillators in Sec II using the RG approach. The quasi-periodic frequencies and the fixed amplitudes of the modes are determined in terms of the system parameters. Numerical support is provided for the existence of the "fixed torus" in phase space. In Sec III, we turn to the Kuznetsov oscillator \cite{Kuznetsov_2010}. As far as we can tell, this was the first investigation (this work was numerical) of a quasi-periodic orbit in a three dimensional dynamical system. We provide in this section a RG analysis (coupled with an adiabatic approximation) of the Kuznetsov oscillator predicting the fixed amplitudes of the two modes in terms of the system parameters. The parameter dependence of the amplitudes is verified numerically. In Sec IV we study two coupled forced Van der Pol oscillators with a view to look for entrainment conditions. Since we are looking for a periodic orbit in this case, it suffices to use the Krylov -Bogoliubov technique. The interesting result that we obtain is the possibility of development of entrainment when the two independent oscillators (before coupling) are not individually entrained. We conclude with a short summary in Sec V.

\section{Coupled Van der Pol oscillator}\label{sec:cvpo}

In this section, we consider the existence of a fixed attractor for quasiperiodic orbits in a set of coupled non identical Van der Pol oscillator which attract all initial conditions and consequently is a higher dimensional generalization of a limit cycle found in a single Van der Pol oscillator. We will see that the renormalization group treatment is ideally suited for obtaining such an attractor. We will provide numerical support for the existence of the attractor. The coupled Van der Pol oscillator system that we consider can be written as ($\omega_1\neq \omega_2$)
\begin{subequations}
  \begin{align}
    \ddot{x} + k\dot{x}(x^2-1) + \omega_1^2x &=  \alpha y \label{eq:cvpo_a}\\
    \ddot{y} + k\dot{y}(y^2-1) + \omega_1^2y &=  \alpha x. \label{eq:cvpo_b}
  \end{align}
\end{subequations}
In the above $\omega_1,\omega_2,\alpha$ and $k$ are constants. We consider $k$ and $\alpha$ to be small so that we can develop a perturbation theory jointly in $k$ and $\alpha$. The ratio $\omega_1/\omega_2$ need not be a rational number. Accordingly, we expand
\begin{subequations}
  \begin{align}
    x &= x_{00} + k x_{10} + \alpha x_{01} + k^2 x_{20} + \alpha kx_{11} + \alpha^2 x_{02} + \hdots \\
    y &= y_{00} + k y_{10} + \alpha y_{01} + k^2 y_{20} + \alpha ky_{11} + \alpha^2 y_{02} + \hdots 
  \end{align}
  \label{eq:expansionofvariables}
\end{subequations}

Inserting the above expansions in Eqs.~(\ref{eq:cvpo_a}) and (\ref{eq:cvpo_b}) and equating terms with the same powers of $k^m \alpha^n$ ($m,n = 0,1,2,\hdots$), we get at different orders, for $x_{mn}(t)$
\begin{subequations}
  \begin{align}
    \ddot{x}_{00} + \omega_1^2x_{00} &= 0 \label{eq:DiffEqn_a}\\
    \ddot{x}_{10} + \omega_1^2x_{10} &= -\dot{x}_{00}(x_{00}^2-1) \\
    \ddot{x}_{01} + \omega_1^2x_{01} &= y_{00} \label{eq:DiffEqn_c}\\
    \ddot{x}_{20} + \omega_1^2x_{20} &= \dot{x}_{10}- \frac{\rm{d}}{\rm{dt}}(x_{00}^2x_{10})\\
    \ddot{x}_{11} + \omega_1^2x_{11} &= -\dot{x}_{01}(x_{00}^2-1) - x_{01}\frac{\rm{d}}{\rm{dt}}x_{00}^2 + y_{10}\\
    \ddot{x}_{02} + \omega_1^2x_{02} &= y_{01}
  \end{align}
\end{subequations}

For the corresponding equations for $y$ we simply replace $\omega_1$ by $\omega_2$, $x_{mn}$ by $y_{mn}$ and vice versa.

Fixing the initial conditions at time $t=t_0$ and choosing $x(t_0) = A(t_0)$, $y(t_0) = B(t_0)$ and $\dot{x}(t_0)=\dot{y}(t_0) = 0$, we write
\begin{align}
  \begin{split}
    x_{00} = A(t_0)\cos \omega_1(t-t_0) \\
    y_{00} = B(t_0)\cos \omega_2(t-t_0) 
  \end{split}
\end{align}
This leads to
\begin{align}
  \ddot{x}_{10} + \omega_1^2x_{10} &= A\omega_1\Big[ (\frac{A^2}{4}-1)\sin \omega_1(t-t_0) \nonumber \\ &\hspace{1cm} + \frac{A^2}{4} \sin 3\omega_1(t-t_0) \Big] \\
  \ddot{x}_{01} + \omega_1^2x_{01} &= B(t_0)\cos \omega_2(t-t_0) 
\end{align}

The solution for $x(t)$ can be written down to first order as
\begin{align}
  x(t) = &A(t_0)\cos\omega_1(t-t_0) \nonumber \\ & - kA\Big[ (\frac{A^2}{4}-1)\frac{(t-t_0)}{2}\cos\omega_1(t-t_0) \nonumber \\ & \hspace{1cm}+ \frac{A^2}{32\omega_1}\sin 3\omega_1(t-t_0) \Big] +\alpha \frac{B(t_0)}{\omega_1^2-\omega_2^2}\cos\omega_2(t-t_0) \nonumber \\ & + \rm{higher~order~terms~in~}\alpha~\rm{and}~k \label{eq:baresolution}
\end{align}

The secular term $(t-t_0)$ needs to be removed to obtain a finite perturbation and at this point we introduce the renormalization constants $Z(t_0,\tau)$ for amplitude and $\tilde{Z}(t_0,\tau)$ for phase as follows
\begin{subequations}
  \begin{align}
    A(t_0) &= Z(t_0,\tau)A(\tau) \label{eq:RGdefinition_a}\\
    -\omega_1t_0 &=\theta(\tau)+\tilde{Z}(t_0,\tau) \label{eq:RGdefinition_b}\\
    B(t_0) &= Z'(t_0,\tau)B(\tau) \label{eq:RGdefinition_c}\\
    -\omega_2t_0 &=\theta'(\tau)+\tilde{Z}'(t_0,\tau) \label{eq:RGdefinition_d}
  \end{align}
\end{subequations}

We expand the renormalization constants as
\begin{subequations}
  \begin{align}
    Z(t_0,\tau) &= 1 + kZ_{10}+\alpha Z_{01} + k^2Z_{20} +\alpha k Z_{11}+\alpha^2Z_{02}+\hdots \\
    \tilde{Z}(t_0,\tau) &= k\tilde{Z}_{10}+\alpha \tilde{Z}_{01} + k^2\tilde{Z}_{20} +\alpha k \tilde{Z}_{11}+\alpha^2\tilde{Z}_{02}+\hdots 
  \end{align}
\end{subequations}

Inserting the above expansions along with the relations shown in Eqs.~(\ref{eq:RGdefinition_a})-(\ref{eq:RGdefinition_d}) into Eq.~(\ref{eq:baresolution}) yields
\begin{align}
  x = &A\cos(\omega_1t+\theta) + kZ_{10}\cos(\omega_1t+\theta) \nonumber \\ & + \alpha Z_{01}\cos(\omega_1t+\theta) - A\sin(\omega_1t+\theta)[k\tilde{Z}_{10}+\alpha\tilde{Z}_{01}]\nonumber \\ & - k\frac{A}{2}(\frac{A^2}{4}-1)(t-\tau+\tau-t_0)\cos(\omega_1t+\theta) \nonumber \\ &   + \alpha\frac{B}{\omega_1^2-\omega_2^2}\cos(\omega_2t+\theta')-\frac{kA^3}{32\omega_1}\cos 3(\omega_1t+\theta)
\end{align}

The removal of the secular term at $\mathcal{O}(k)$ yields
\begin{align}
  \begin{split}
    Z_{10}&=-\frac{A}{2}(\frac{A^2}{4}-1)(\tau-t_0) \\
    Z_{01}&= \tilde{Z}_{10}=\tilde{Z}_{01}=0
  \end{split}
\end{align}
We require $x(t,\tau)$(?) to be independent of $\tau$, as it must be since $x(t)$ is independent of where the initial condition is on the trajectory. This implies $\frac{\rm{d}}{\rm{d}\tau}x(t,\tau)=0$ and leads to (after identifying $\tau$ as the current time $t$)
\begin{align}
  \frac{\rm{d}A}{\rm{d}t} = -k\frac{A}{2}(\frac{A^2}{4}-1). \label{eq:RGequations_A}
\end{align}
An identical treatment of the analogues of Eqs.~(\ref{eq:DiffEqn_a})-(\ref{eq:DiffEqn_c}) for $y(t)$ leads to the flow
\begin{align}
  \frac{\rm{d}B}{\rm{d}t} = -k\frac{B}{2}(\frac{B^2}{4}-1). \label{eq:RGequations_B}
\end{align}
At this stage (i.e., after carrying out $\mathcal{O}(\alpha)$ and $\mathcal{O}(k)$ calculations) our solution reads
\begin{subequations}
  \begin{align}
    x_{00} &= A\cos(\omega_1t+\theta) \\
    x_{10} &= -\frac{A^3}{32\omega_1}\sin 3(\omega_1t+\theta)\\
    x_{01} &= \frac{B}{\omega_1^2-\omega_2^2}\cos (\omega_2t+\theta')\\
    y_{00} &= B\cos(\omega_2t+\theta') \\
    y_{10} &= -\frac{B^3}{32\omega_2}\sin 3(\omega_2t+\theta')\\
    y_{01} &= \frac{A}{\omega_2^2-\omega_1^2}\cos (\omega_1t+\theta)
  \end{align}
\end{subequations}
with the coefficients $A(t)$ and $B(t)$ given by the flow equationsEqs.~(\ref{eq:RGequations_A}) and (\ref{eq:RGequations_B}). 

We now proceed to the next order of calculation, the details of which can be found in the Appendix. At this second order of perturbation theory, there is no correction to Eqs.~(\ref{eq:RGequations_A}) and (\ref{eq:RGequations_B}), but the phases $\theta$ and $\theta'$ evolve according to
\begin{subequations}
  \begin{align}
    \frac{\rm{d}\theta}{\rm{dt}} &= -\frac{\alpha^2}{2\omega_1}\frac{1}{\omega_2^2-\omega_1^2}+\frac{k^2A^4}{256\omega_1} \label{eq:RGequations_th}\\
        \frac{\rm{d}\theta'}{\rm{dt}} &= -\frac{\alpha^2}{2\omega_2}\frac{1}{\omega_1^2-\omega_2^2}+\frac{k^2B^4}{256\omega_2} \label{eq:RGequations_thp}
  \end{align}
\end{subequations}

Since we are interested in finding the asymptotic solution (i.e., the solution for $t\rightarrow \infty$), we note that both $A$ and $B$ tend to 2 with no correction at $\mathcal{O}(k^2)$ in perturbation theory. Hence to this order we can replace $A$ and $B$ by their asymptotic value of 2. With these substitutions in Eqs.~(\ref{eq:RGequations_th}) and (\ref{eq:RGequations_thp}), we find that the frequencies $\omega_1$ and $\omega_2$ are renormalized to
\begin{subequations}
  \begin{align}
    \omega_1'^{2} &= \omega_1^2-\frac{\alpha^2}{2\omega_1}\frac{1}{\omega_2^2-\omega_1^2}+\frac{k^2}{16\omega_1} \label{eq:omega_p_a} \\
    \omega_2'^{2} &= \omega_2^2-\frac{\alpha^2}{2\omega_2}\frac{1}{\omega_1^2-\omega_2^2}+\frac{k^2}{16\omega_2} \label{eq:omega_p_b} .
  \end{align}
\end{subequations}

The amplitudes $A$ and $B$ of `primary' oscillation of $x$ and $y$ are seen to be $2+\mathcal{O}(k^2)$ and independent of $\alpha$ upto $\mathcal{O}(\alpha^2)$.

Having noted that for $k\ll 1$, periodic orbits of amplitude $2$ are formed independent of $\alpha$, we drop the $k$-terms in Eqs.~(\ref{eq:cvpo_a}) and (\ref{eq:cvpo_b}) and find that resulting coupled set of differential equations have the eigenvalue
\begin{align}
  \Omega_{1,2}^2 = \frac{1}{2}\Big[ \omega_1^2 + \omega_2^2 \pm \sqrt{(\omega_1^2-\omega_2^2)^2+4\alpha^2} \Big]
\end{align}
Expanding upto $\mathcal{O}(\alpha^2)$, we find precisely the frequencies $\omega_1'^2$ and $\omega_2'^2$ shown in Eqs.~(\ref{eq:omega_p_a}) and (\ref{eq:omega_p_b}), when $\mathcal{O}(k^2)$ terms are ignored. This allows us to make the statement that for $k\ll 1$, the coupled equations shown in Eqs.~(\ref{eq:cvpo_a}) and (\ref{eq:cvpo_b}) have the solution
\begin{align}
  \begin{split}
    x(t)&=2\cos\Omega_1t+\frac{2\alpha}{\Omega_1^2-\Omega_2^2}\cos\Omega_2t - \frac{k}{4\Omega_1}\sin 3\Omega_1t+\hdots\\
    y(t)&=2\cos\Omega_2t+\frac{2\alpha}{\Omega_2^2-\Omega_1^2}\cos\Omega_1t - \frac{k}{4\Omega_2}\sin 3\Omega_2t+\hdots
  \end{split}\label{eq:RG_solution_cvpo}
\end{align}

Choosing $\Omega_1/\Omega_2$ as an irrational numbers, we have a quasiperiodic orbit winding around a torus of definite radii in two perpendicular planes.

For small $\alpha$, we make a transformation to the variables $X = x + \frac{\alpha}{\Omega_2^2-\Omega_1^2}y$ and $Y = y - \frac{\alpha}{\Omega_2^2-\Omega_1^2}x$ and find
\begin{align}
  X =& (2+\frac{2\alpha^2}{(\Omega_2^2-\Omega_1^2)^2})\cos\Omega_1t +  \mathcal{O}(k) \nonumber \\
  Y =& (2+\frac{2\alpha^2}{(\Omega_2^2-\Omega_1^2)^2})\cos\Omega_2t +  \mathcal{O}(k)
       \label{eq:torus_solution}
\end{align}

For $\Omega_1/\Omega_2$ an irrational number, the existence of a fixed torus in $X,\dot{X},Y,\dot{Y}$ space is obvious from Eq. (\ref{eq:torus_solution}).

A different kind of coupled Van der Pol oscillator which can give a 'fixed' torus is the following variation on a model studied by \cite{Pastor_1993}. We consider
\begin{align}
  &\ddot{x} + k\dot{x}(x^2+\alpha y^2-1) + \omega_1^2x = 0 \nonumber\\
  &\ddot{y} + k\dot{y}(y^2+\alpha x^2-1) + \omega_2^2y = 0 
\end{align}
where $\omega_1/\omega_2$ is an irrational numbers. As before, $k$ and $\alpha$ are much smaller than unity and we carry out a double power series expansion exactly as in Eqs. (\ref{eq:expansionofvariables}a) and (\ref{eq:expansionofvariables}b). At the zeroth order, we get
\begin{align}
  x_{00} = A\cos\omega_1(t-t_0)\nonumber\\
  y_{00} = B\cos\omega_2(t-t_0).
\end{align}

At $\mathcal{O}(k)$,
\begin{align}
  \ddot{x}_{10} + \omega_1^2x_{10}=-\dot{x}_{00}(x^2_{00}-1) \nonumber\\
  \ddot{y}_{10} + \omega_2^2y_{10}=-\dot{y}_{00}(y^2_{00}-1).
\end{align}

As before this leads to the flow equations (see Eqs. (\ref{eq:RGequations_A}) and (\ref{eq:RGequations_B}))
\begin{align}
  \frac{\text{d}}{\text{dt}}A = -k\frac{A}{2}(\frac{A^2}{4}-1) \nonumber\\
  \frac{\text{d}}{\text{dt}}B = -k\frac{B}{2}(\frac{B^2}{4}-1)
\end{align}
and
\begin{align}
  x_{10}=-\frac{A^3}{32\omega_1}\sin 3(\omega_1t+\theta) \nonumber\\
  y_{10}=-\frac{B^3}{32\omega_2}\sin 3(\omega_2t+\theta) \nonumber\\
\end{align}

At $\mathcal{O}(k^2)$, we pick up a correction to the frequencies $\omega_1$ and $\omega_2$ and at $\mathcal{O}(\alpha^2)$, there is no contribution to $x$ and $y$.

The interesting contribution comes at $\mathcal{O}(\alpha k)$, where
\begin{subequations}
  \begin{align}
    \ddot{x}_{11}+\omega^2_1x_{11} &= -\dot{x}_0y_0^2 \nonumber\\ &\hspace{-1cm}=\omega_1\frac{A}{2}\sin \omega_1(t-t_0)[B^2(1+\cos 2\omega_2(t-t_0))] \\
    \ddot{y}_{11}+\omega^2_2y_{11} &= -\dot{y}_0z_0^2\nonumber\\ &\hspace{-1cm}=\omega_2\frac{B}{2}\sin \omega_2(t-t_0)[A^2(1+\cos 2\omega_1(t-t_0))] 
  \end{align}
  \label{eq:dynamicalequationsnew}
\end{subequations}

Removal of the resonance Eq. (\ref{eq:dynamicalequationsnew}a) leads to the flow
\begin{align}
  \frac{\text{d}}{\text{d}\tau}A &=\frac{kA}{2}[1-\frac{A^2}{4}-\alpha\frac{B^2}{2}]
\end{align}
and the removal in Eq. (\ref{eq:dynamicalequationsnew}b) yields
\begin{align}
  \frac{\text{d}}{\text{d}\tau}B &=\frac{kB}{2}[1-\frac{B^2}{4}-\alpha\frac{A^2}{2}].
\end{align}

The fixed points of the above flow equations give
\begin{align}
  1=\frac{A^2}{4}+\frac{\alpha B^2}{2} = \frac{B^2}{4}+\frac{\alpha A^2}{2}. 
\end{align}

This leads to $A^2=B^2$ and hence
\begin{align}
  A^2=B^2=\frac{4}{1+2\alpha}.
\end{align}

At this order, the solution reads
\begin{subequations}
  \begin{align}
    x &= \frac{2}{\sqrt{1+2\alpha}}\cos \omega_1t \nonumber\\
    y &= \frac{2}{\sqrt{1+2\alpha}}\cos \omega_2t .
  \end{align}
  \label{eq:solutionfirstorder}
\end{subequations}

There will be $\mathcal{O}(k^2)$ corrections to the frequencies $\omega_1$ and $\omega_2$ but for $k\ll 1$ (e.g., $k=0.1$), these corrections are negligible. Hence if we choose $\omega_1/\omega_2$ as irrational, then the 'fixed' torus in the $x,\dot{x},y,\dot{y}$ space will be given by Eqs. (\ref{eq:solutionfirstorder}a) and (\ref{eq:solutionfirstorder}b). For $\alpha=1/2$, $k=0.1$ and $\omega_1=2, \omega_2=1\Rightarrow\Omega_{1,2}^2=\frac{1}{2}(5\pm\sqrt{10})$, we show the time series and the torus in Fig. 1. The amplitude of the limit cycle is in accordance with Eqs. (\ref{eq:solutionfirstorder}a) and (\ref{eq:solutionfirstorder}b) for $\alpha=1/4$.

Eq.~(\ref{eq:RG_solution_cvpo}) suggests that the main contributions to the time series $x$ and $y$ happen to come from these two frequencies $\Omega_1$ and $\Omega_2$. Therefore, we can set the ratio of these two frequencies to be irrational so that in the phase-space $(x,~\dot{x}=z,~y,~\dot{y}=w)$, the phase-space orbit from a single initial state fills up the whole higher dimensional torus as $t\longrightarrow\infty$. 
\begin{figure}[h!]
\centering
\includegraphics[scale=0.3]{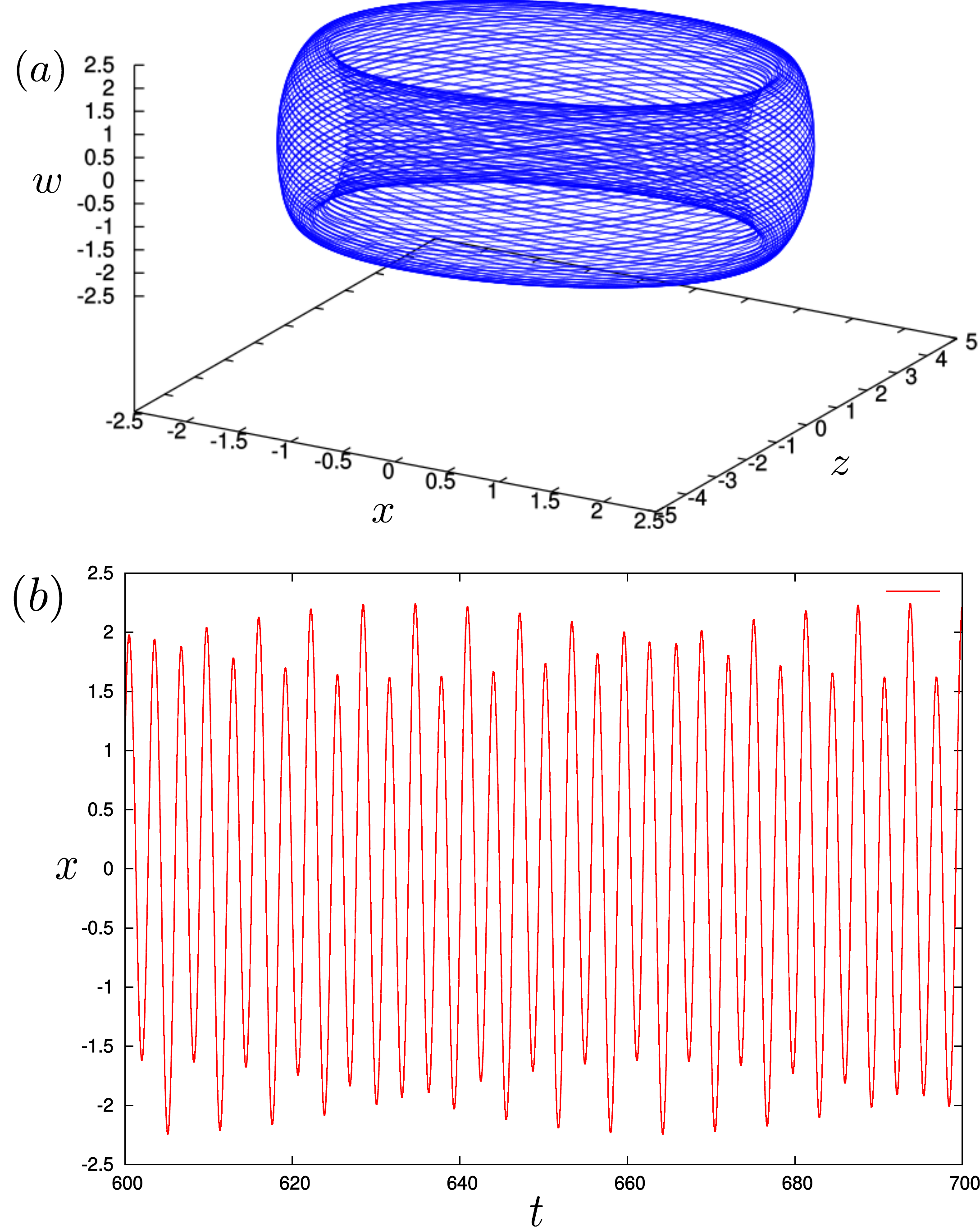}
\caption{Three dimensional projection of the phase space trajectories, torus formed (a) and two frequency quasiperiodic response, $x$ vs $t$ (b)} \label{fig:qperiodic}
\end{figure}

\section{RG analysis of the Kuznetsov Oscillator}\label{sec:qprdc}

In this section, we outline the application of the renormalization group technique in another coupled three-dimensional system. This model was first introduced \cite{Kuznetsov_2010} as an autonomous stable quasiperiodic self-oscillator with a two dimensional torus as the attractor and can be possibly experimentally implemented, for example, as an electronic device.
\begin{align}
   & \ddot{x} - (\lambda + z+x^2-\frac{x^4}{2})\dot{x} + \omega_0^2x = 0 \nonumber \\
    & \dot{z} = \mu - x^2 \label{eq:dynamicalequation}
\end{align} 
In order to solve this equation perturbatively, we multiply the $\dot{x}$ term with a dimensionless parameter $\epsilon$. At the end of the analysis, we want to impose $\epsilon = 1$. Also, we always work in the $\lambda = 0 $ limit. Hence, the above equation becomes:
\begin{equation}
  \ddot{x} - \epsilon (z+x^2-\frac{x^4}{2})\dot{x} + \omega_0^2x = 0. \nonumber
\end{equation}
We expand $x$ using the perturbation parameter: $x = x_0 + \epsilon x_1+\hdots{}$.
\begin{align}
\frac{\rm{d}^2}{\rm{dt}^2}(x_0 + \epsilon x_1)  & - \epsilon \Big[z+(x_0 + \epsilon x_1)^2-\frac{(x_0 + \epsilon x_1)^4}{2}\Big](\dot{x}_0+\epsilon\dot{x}_1) \nonumber \\ &  + \omega_0^2(x_0 + \epsilon x_1) = 0.
\end{align}
We now equate terms to zero order by order in $\epsilon$.

\begin{equation}
  \begin{aligned}
    &\ddot{x}_0 + \omega_0^2x_0 = 0 \nonumber \\
    &\ddot{x}_1 - (z+x_0^2-\frac{x_0^4}{2})\dot{x}_0  + \omega_0^2x_1 = 0 \nonumber
  \end{aligned}
\end{equation}
Now, we try to solve these equations for a fixed $z$ at each order in $\epsilon$, starting with
\begin{equation}
  x_0(t) = C\cos\omega_0t
\end{equation}
where we have used the initial conditions $x(t=0)=\rm{C}$ and $\dot{x}(t=0)=0$. This solution will be used in the $\mathcal{O}(\epsilon)$ equation.
\begin{equation}
  \begin{aligned}
    &\ddot{x}_1 + \omega_0^2x_1 = - \Big[z + C^2\cos^2\omega_0t - {1\over 2}C^4\cos^4{\omega_0t}\Big]\omega_0C \sin \omega_0t  \nonumber \\
    \Rightarrow ~ &\ddot{x}_1 + \omega_0^2x_1 = -\omega_0C\sin \omega_0t\Big[z + \frac{C^2}{2}(\cos 2\omega_0t + 1) \nonumber \\ & \hspace{2.5cm} - {1\over 16}C^4(\cos 4{\omega_0t} + 4\cos 2{\omega_0t} + 3) \Big]  \nonumber \\
    \Rightarrow ~ &\ddot{x}_1 + \omega_0^2x_1 = -\omega_0C \sin \omega_0t \Big[ (z + \frac{C^2}{2} - {3C^4\over 16}) \nonumber \\ & \hspace{2.5cm} +  (\frac{C^2}{2} - {C^4\over 4})\cos 2\omega_0t  - {C^4\over 16} \cos 4{\omega_0t}  \Big] \nonumber   \\
    \Rightarrow ~ &\ddot{x}_1 + \omega_0^2x_1 = -\tilde{\alpha} \sin \omega_0t - \tilde{\beta} \sin \omega_0t \cos 2 \omega_0t \nonumber \\ & \hspace{2.5cm} - \tilde{\gamma} \sin \omega_0t \cos 4 \omega_0t \nonumber   
  \end{aligned}
\end{equation}
where $\tilde{\alpha} = \omega_0C(z+{C^2\over 2}-{3C^4\over 16})$, $\tilde{\beta} = \omega_0C({C^2\over 2}-{C^4\over 4})$ and $\tilde{\gamma} = -{\omega_0C^5\over 16}$. Using some identities, the final equation is
\begin{equation}
  \ddot{x}_1+\omega_0^2x_1 = -\alpha\sin\omega_0t -\beta\sin 3\omega_0t -\gamma\sin 5\omega_0t
\end{equation}
where $\alpha = \tilde{\alpha} - {\tilde{\beta}\over 2}$, $\beta = \frac{\tilde{\beta} - \tilde{\gamma}}{2}$ and $\gamma = {\tilde{\gamma}\over 2}$.

The solution can be written as
\begin{align}
  x_1(t) =& {\alpha t\over 2\omega_0}\cos\omega_0t + {\beta \over 8\omega_0^2}\sin 3\omega_0t + {\gamma \over 24\omega_0^2}\sin 5\omega_0t \nonumber \\ & \hspace{2.5cm} + A\sin\omega_0t+B\cos\omega_0t 
\end{align}
where $A$ and $B$ are fixed by the boundary conditions. It is to be noted that the first term diverges as $t\rightarrow \infty$ and is unphysical. This term needs to be treated specially and the divergence has to be removed.

Now, $x_1(t=0) = 0 \Rightarrow B = 0$ and $\dot{x}_1(t=0) = 0 \Rightarrow A = -\Big({\alpha\over 2\omega_0^2} + {3\beta\over 8\omega_0^2} + {5\gamma\over 24\omega_0^2} \Big)$.

Hence, upto $\mathcal{O}(\epsilon)$, the full solution is
  \begin{align}
    x(t) &= x_0+\epsilon x_1  \nonumber \\
    &= C\cos\omega_0t + \epsilon\Big( {\alpha t\over 2\omega_0}\cos\omega_0t + {\beta \over 8\omega_0^2}\sin 3\omega_0t \nonumber \\ & \hspace{1.5cm} + {\gamma \over 24\omega_0^2}\sin 5\omega_0t + A\sin\omega_0t \Big) \nonumber \\
    &= C\cos\omega_0t + \epsilon\Big( {\alpha (t-\tau)\over 2\omega_0}\cos\omega_0t + {\beta \over 8\omega_0^2}\sin 3\omega_0t \nonumber \\ & \hspace{1.5cm} + {\gamma \over 24\omega_0^2}\sin 5\omega_0t + A\sin\omega_0t \Big) + \epsilon {\alpha \tau\over 2\omega_0}\cos\omega_0t \label{eq:fullsolution}
  \end{align}
where we have introduced an arbitrary time $\tau$ to split the full time interval into $t-\tau$ and $\tau-0=\tau$. The amplitude and the phase of the solution now has dependence on $\tau$ due to non-linearities of the system. The terms containing $\tau-0$ needs to be now absorbed in the renormalized parts of the amplitude and the phase.

We perform a expansion of the amplitude and the phase in $\epsilon$.

\begin{center}
  \centering
  \begin{figure}[t]
    \includegraphics[scale=0.6]{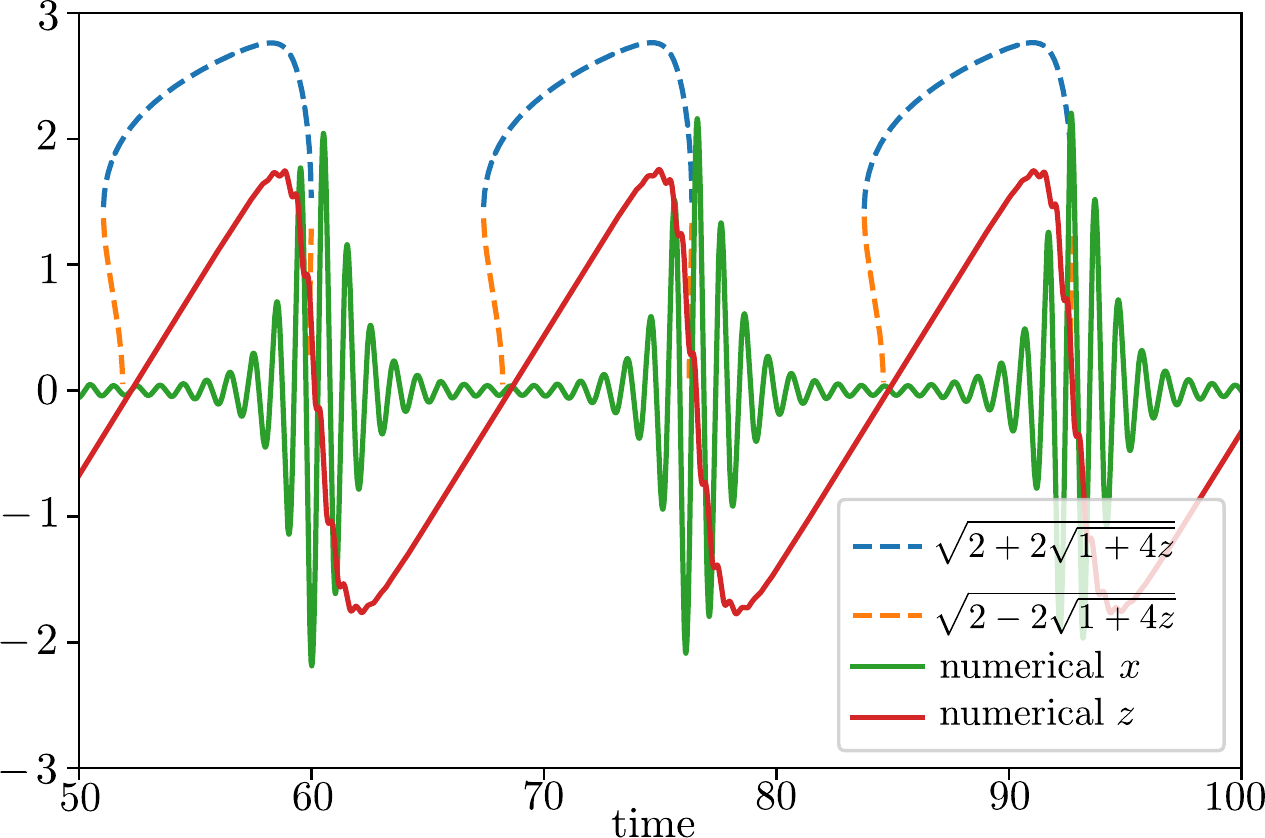}
    \caption{Location of the limit cycle points as a function of numerically obtained $z$. The parameters used here are $\omega_0= 2\pi, \mu = 0.3, \lambda = 0$.}
    \label{fig:limitcycle}
  \end{figure}
\end{center}

\begin{figure*}[t]
  \includegraphics[scale=0.95]{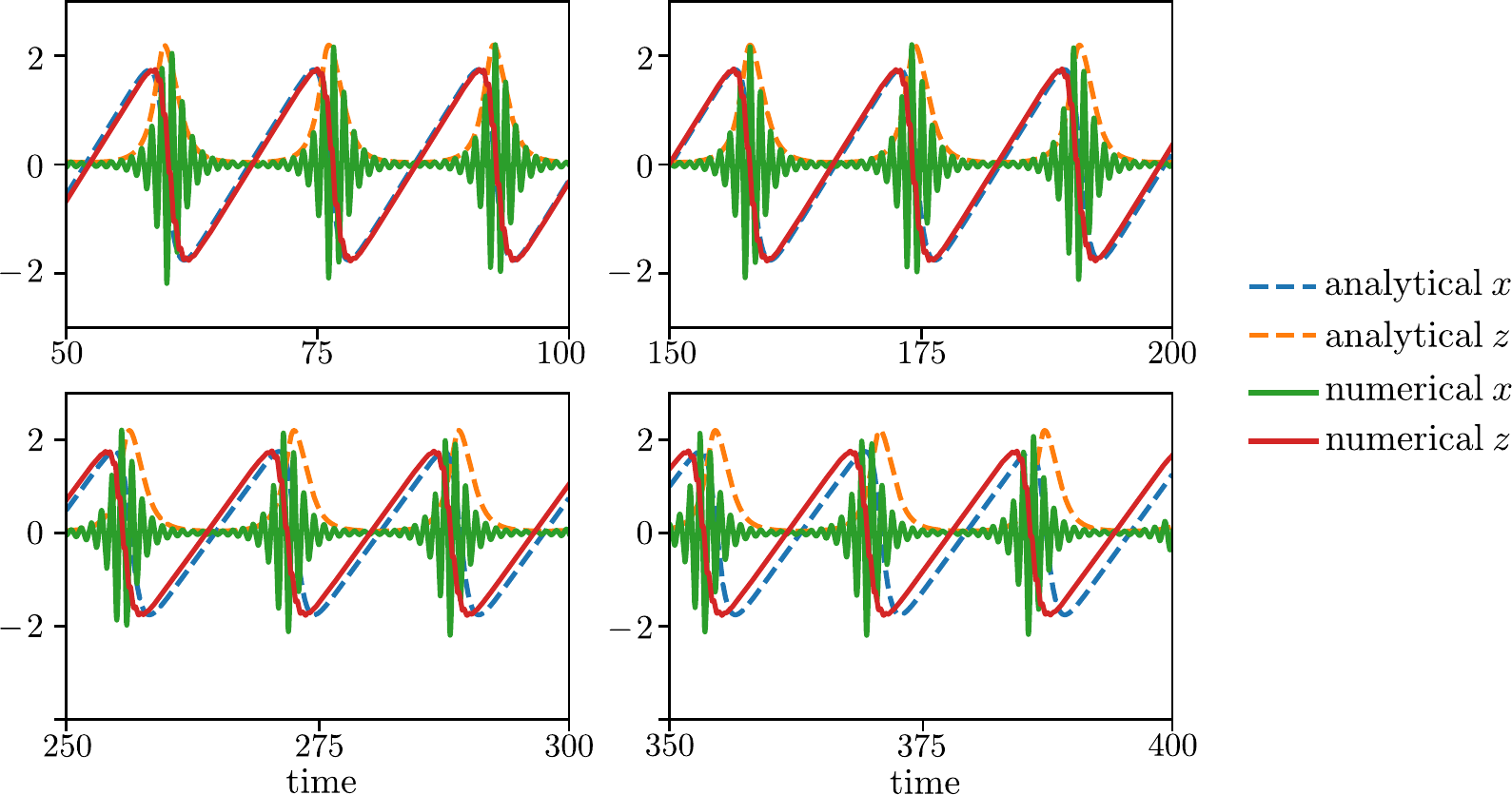}
  \caption{Comparison of the numerical solutions of Eq.~(\ref{eq:dynamicalequation}) given by the continuous line and Eq.~(\ref{eq:RGequations}) given by the dashed line. The parameters used here are the same as Fig. \ref{fig:limitcycle}.}
  \label{fig:fullcomparison}
\end{figure*}

  \begin{figure}[b]
    \includegraphics[scale=0.9]{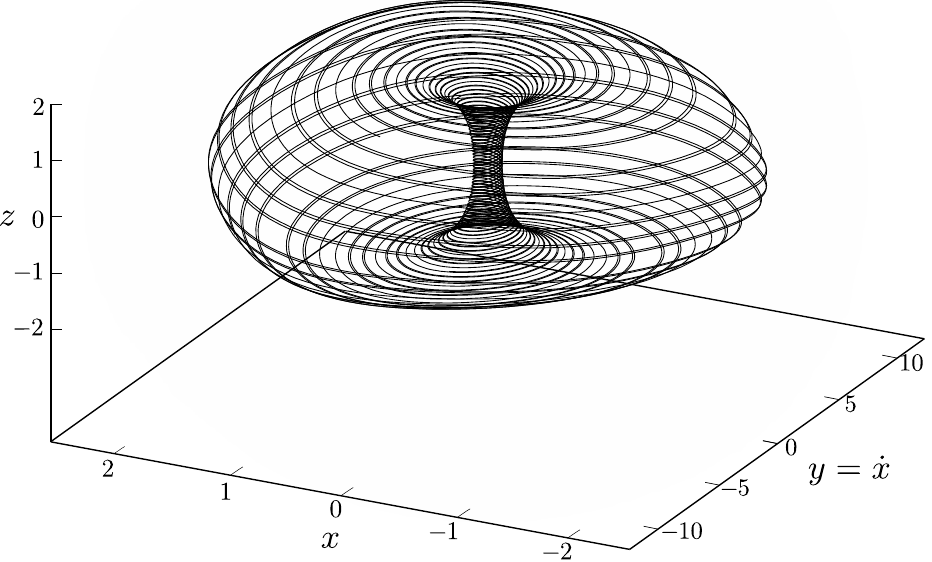}
    \caption{Phase space explored by the Kuznetsov oscillator lies on a torus. The trajectories get denser and denser and eventually fills the entire torus at long times. The parameters used here are $\omega_0= 2\pi, \mu = 0.4, \lambda = 0$.}
    \label{fig:kuznetsov_torus}
  \end{figure}

\begin{align}
\begin{split}
  C &= C_\tau(1+a_1\epsilon+ \hdots) \\
    0 &= \theta_\tau + a_2\epsilon + \hdots \label{eq:amplitudeexpansion}
\end{split}
  \end{align}

since the initial phase was $\theta(t=0)=0$. Now, substituting Eq. (\ref{eq:amplitudeexpansion}) into Eq. (\ref{eq:fullsolution}), it is easy to see that only the $x_0$ solution contributes to the expansion of the amplitude as we are retaining terms only upto $\mathcal{O}(\epsilon)$. This expansion is
\begin{align}
  C\cos\omega_0t &= C_\tau(1+a_1\epsilon)\cos(\omega_0t + \theta_\tau + a_2\epsilon) \nonumber \\
  &=  C_\tau\cos(\omega_0t + \theta_\tau) - \epsilon C_\tau a_2\sin(\omega_0t + \theta_\tau) \nonumber \\ & \hspace{3cm} + \epsilon a_1\cos(\omega_0t + \theta_\tau). 
\end{align}
In the rest of the terms, we can consistently substitute $C ~\rm{by}~ C_\tau$ and $\omega_0t ~\rm{by}~(\omega_0t+\theta_\tau)$ upto $\mathcal{O}(\epsilon)$. It is easy to see that in order to cancel the diverging term, we have to choose $a_2=0$ and $a_1=-{\alpha\tau\over 2\omega_0}$.

Hence,
\begin{align}
  x(t) = &C_\tau\cos(\omega_0t+\theta_\tau) + \epsilon\Big( {\alpha_\tau (t-\tau)\over 2\omega_0}\cos(\omega_0t+\theta_\tau) \nonumber \\ & \hspace{0cm} + {\beta_\tau \over 8\omega_0^2}\sin (3\omega_0t+\theta_\tau) + {\gamma_\tau \over 24\omega_0^2}\sin (5\omega_0t+\theta_\tau)\nonumber\\ &+ A_\tau\sin(\omega_0t+\theta_\tau) \Big)
\end{align}
where the subscript $\tau$ in $\alpha_\tau, \beta_\tau, \gamma_\tau$ and $A_\tau$ signifies that all the functional dependence of these terms on $C$ is now replaced by $C_\tau$.

Now, since $\tau$ is arbitrarily chosen, $\frac{\rm{d}}{\rm{d}\tau}x(t)=0$. From this, we can equate the coefficients of $\cos\omega_0t$ and $\sin\omega_0t$ to be independently equal to zero. This gives us the following two equations:
\begin{align}
  \rm{Coefficient~of~}\cos\omega_0t &\Rightarrow \frac{\rm{d}C_\tau}{\rm{d}\tau} -\epsilon{\alpha_\tau\over 2\omega_0} + \epsilon A_\tau \frac{\rm{d}\theta_\tau}{\rm{d}\tau} = 0 \nonumber \\
    \rm{Coefficient~of~}\sin\omega_0t &\Rightarrow -C_\tau \frac{\rm{d}\theta_\tau}{\rm{d}\tau} + \epsilon  \frac{\rm{d}A_\tau}{\rm{d}\tau} = 0
\end{align}

Now, $A_\tau$ is a polynomial of $C_\tau$ and hence, $\frac{\rm{d}A_\tau}{\rm{d}\tau} \propto \frac{\rm{d}C_\tau}{\rm{d}\tau} \sim \mathcal{O}(\epsilon)$. Using the second equation, it is easy to see that $\frac{\rm{d}\theta_\tau}{\rm{d}\tau} \sim \mathcal{O}(\epsilon^2)$ and can be dropped in our analysis. Hence, the two equations can be written as
\begin{align}
  \frac{\rm{d}C_\tau}{\rm{d}\tau} &= \epsilon{\alpha_\tau\over 2\omega_0} = {\epsilon\over 2}C_\tau \Big(z+{C_\tau^2\over 2} - {C_\tau^4\over 16}\Big) \nonumber \\
  \frac{\rm{d}\theta_\tau}{\rm{d}\tau} &= 0.
\end{align}

Thus, we end up at a limit cycle whenever $\alpha_\tau = 0$. This has three real solutions $C_\tau^0 = 0, C_\tau^+ = \sqrt{2+2\sqrt{1+4z}}, C_\tau^- = \sqrt{2-2\sqrt{1+4z}}$. Using linear stability analysis, it can be seen that $C_\tau^0$ is a stable fixed point for $z<0$ and is unstable for $z>0$. Both the $C_\tau^\pm$ vanish when $z<-0.25$ and only $C_\tau^+$ is present for large positive $z$. Whenever $z>0$ the amplitude tries to grow and approach $C_\tau^+$. When $z<0$, the amplitude tries to decrease and reach the stable fixed point $C_\tau^0$. This is clearly demonstrated in Fig. \ref{fig:limitcycle}.

Now we shift our focus to the dynamical equation of $z$. Upto this point, we have assumed that $z$ remains constant. Now, in the dynamical equation of $z$, we substitute $x^2$ by its average over a cycle: ${C_\tau^2\over 2}$. So, after identifying $\tau = t$, the new set of dynamical equations are written as
\begin{align}
  \dot{C}_t & = {\epsilon\over 2}C_t \Big(z+{C_t^2\over 2} - {C_t^4\over 16}\Big) \nonumber \\
  \dot{z} &= \mu - {C_t^2\over 2} \label{eq:RGequations}
\end{align}

This makes it clear that $z$ increases when $C_t<\sqrt{2\mu}$ and starts to decrease after $C_t$ becomes larger than $\sqrt{2\mu}$. We next wish to compare the solutions of Eq. (\ref{eq:RGequations}) with those of Eq. (\ref{eq:dynamicalequation}). The comparison is shown in Figure \ref{fig:fullcomparison}. The dashed lines are the solutions of the RG equations, Eq. (\ref{eq:RGequations}) and the continuous lines are the solutions of Eq. (\ref{eq:dynamicalequation}). One can see that the solutions of $z$ from both equations have a very large overlap for a very long time. Also, the $C_t$ provides an envelope to the solution of $x$, thereby validating our analysis.

The analysis in this section provides a deeper understanding of the phenomenon demonstrated in Ref. (\cite{Kuznetsov_2010}). In fact, further analysis shows that the solution is periodic (instead of it being quasiperiodic) for smaller values of $\mu$. However, it is not included in this paper. These results clearly demonstrate the applicability of the renormalization group technique in analysing dynamical systems.


\section{Coupled Van der Pol Oscillators Under periodic forcing}
In this section, we return to the coupled Van der Pol oscillators of Equations \eqref{eq:cvpo_a} and \eqref{eq:cvpo_b} with the
difference that they now are now both driven by an external periodic forcing\cite{Ohsuga1985,Choubey2010}. The dynamics is now
governed by
\begin{subequations}
  \begin{align}
  & \ddot{x}+k\dot{x}(x^2-1)+\omega_1^2x=\alpha y+Fcos(\omega_3 t) \label{qE}\\
   & \ddot{y}+k\dot{y}(y^2-1)+\omega_2^2y=\alpha x+Gcos(\omega_3 t) \label{qE_b}
  \end{align}
\end{subequations}
where $\alpha\neq 0$ and $k$ is smaller than unity. We assume the external in-phase periodic forcings to have same frequency $\omega_3$.\\
 We are interested in studying trajectories exhibiting entrainment. Such trajectories are the stable limit cycles
of the system having the same frequency as that of the external forcing. Since there is only one primary frequency scale (that of the external forcing) with the natural frequencies ($\omega_1$ and $\omega_2$) being close to the external frequencies, it is appropriate to use the Krylov-Bogoliubov technique to obtain the flow equation. Accordingly we try the solution
\begin{subequations}
  \begin{align}
    x =& A_1\cos\omega_3 t+ B_1\sin\omega_3 t \\
    y =& A_2\cos\omega_3 t+ B_2\sin\omega_3 t 
  \end{align}
\end{subequations}

Treating $A_{1,2}$ and $B_{1,2}$ as slowly varying quantities, the flow equations are ($k\ll 1$)
\begin{subequations}
  \begin{align}
& 2\omega_3 \dot{A}_1=(\omega_1^2-\omega_3^2) B_1-\alpha B_2+k\omega_3\left(1-\frac{A_1^2+B_1^2}{4}\right)A_1\\
& 2\omega_3 \dot{B}_1=F - (\omega_1^2-\omega_3^2)A_1+\alpha A_2 + k\omega_3\left(1-\frac{A_1^2+B_1^2}{4}\right)B_1 \\
& 2\omega_3 \dot{A}_2= (\omega_2^2-\omega_3^2)B_2-\alpha B_1 + k\omega_3\left(1-\frac{A_2^2+B_2^2}{4}\right)A_2 \\
& 2\omega_3 \dot{B}_2= G - (\omega_2^2-\omega_3^2)A_2+\alpha A_1 + k\omega_3\left(1-\frac{A_2^2+B_2^2}{4}\right)B_2 
  \end{align}
\end{subequations}

\begin{figure}[hbtp]
	\centering
\includegraphics[scale=0.35]{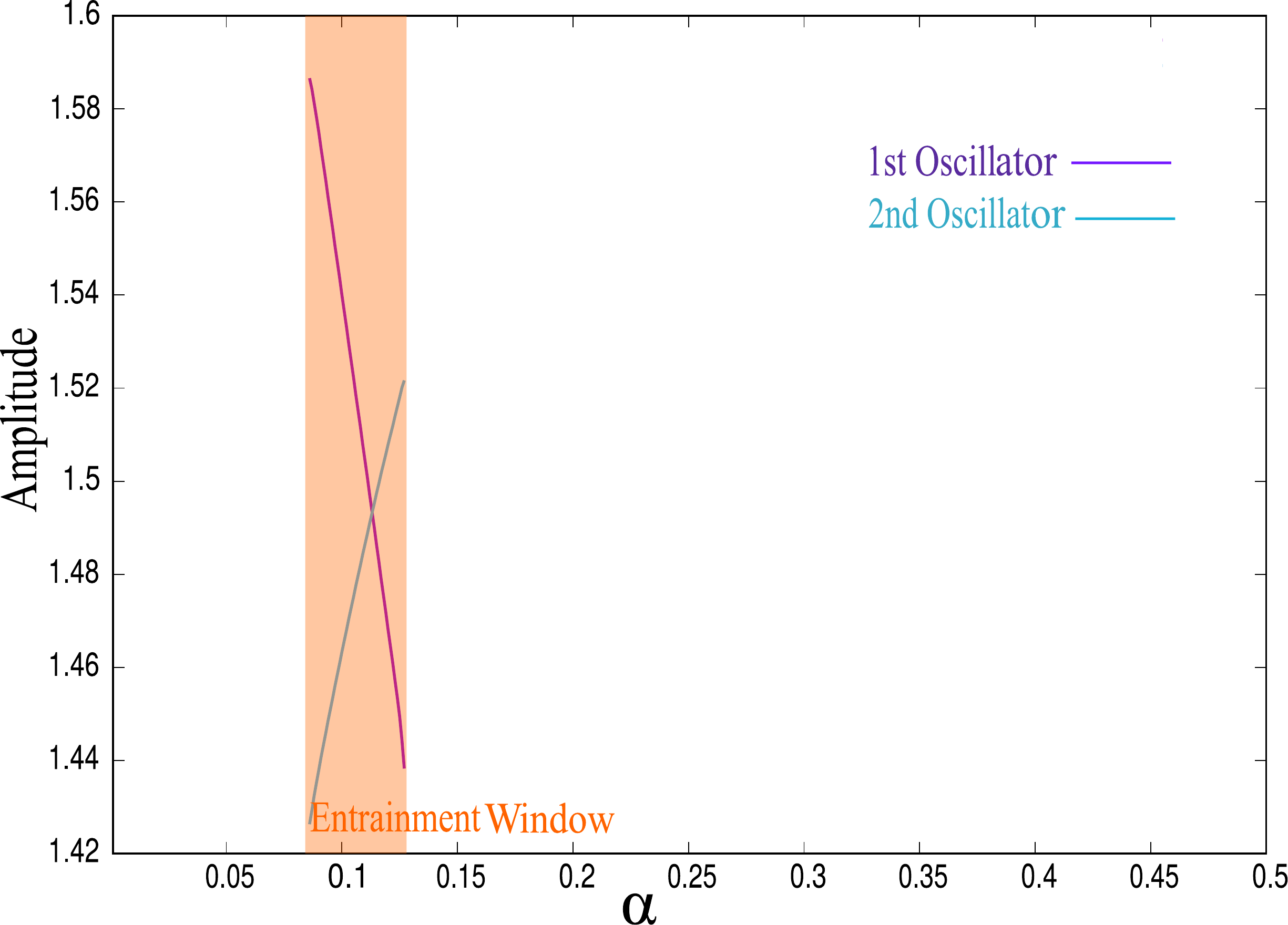}
\caption{The intersection of the shaded region with the $\alpha$ axis gives the range in $\alpha$ such that in that window of $\alpha$, the system is entrained at the frequency of the external periodic forcing. Amplitude of each oscillator is plotted in the entrainment window in coupling constant. The parameters of the system are chosen like this: $\omega_1=4.0729,~\omega_2=3.93084,~\omega_3=4.0,~k=0.1,~F=1.08,~G=0.7$. The entrainment window in $\alpha$ is [0.086:0.127].}
\label{figew}
\end{figure}

\begin{figure}[hbtp]
	\centering
	\includegraphics[scale=0.35]{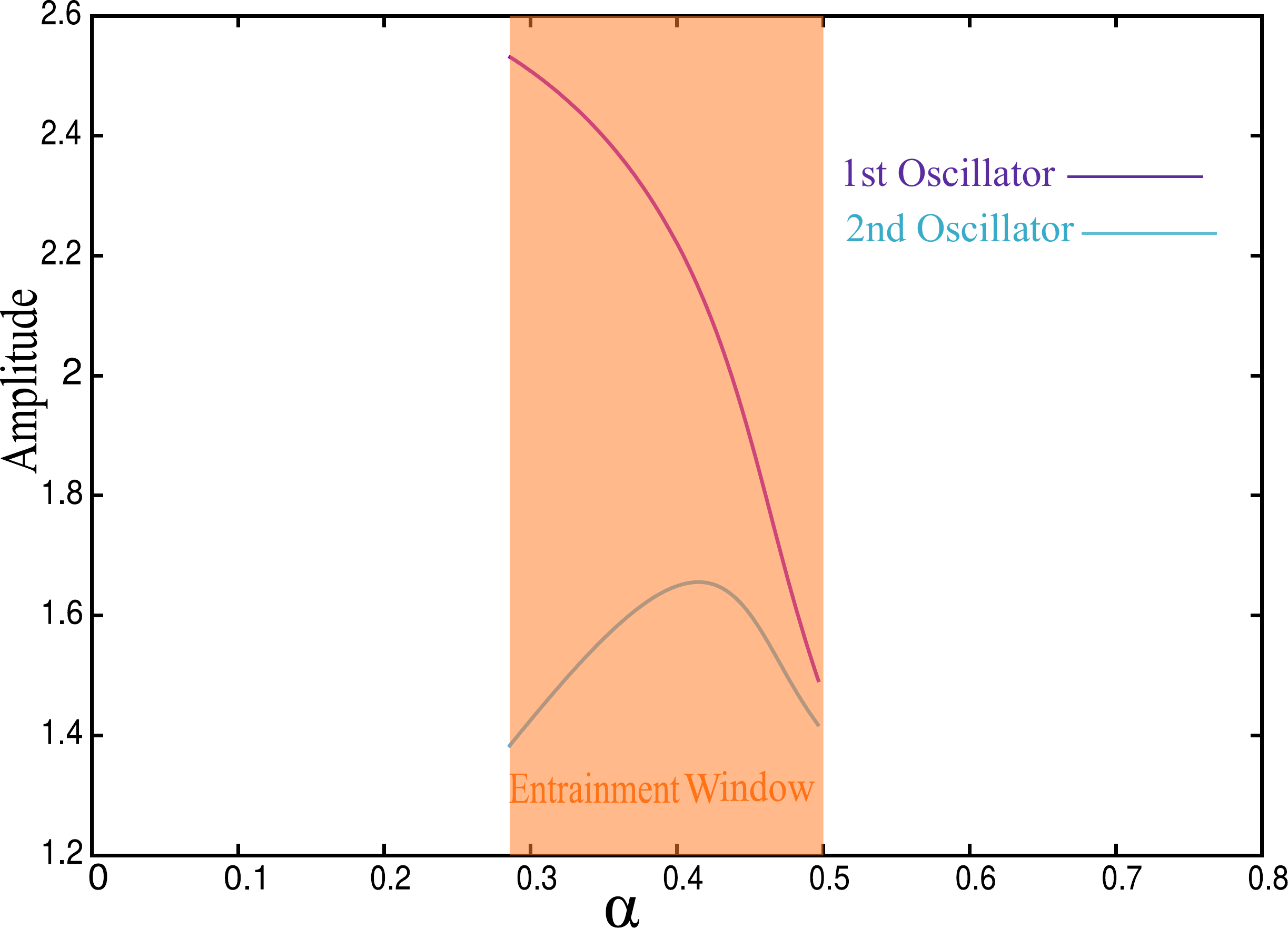}
	\caption{The intersection of the shaded region with the $\alpha$ axis gives the range in $\alpha$ such that in that window of $\alpha$, the system is entrained at the frequency of the external periodic forcing. Amplitude of each oscillator is plotted in the entrainment window in coupling constant. The parameters of the system are chosen like this: $\omega_1=4.0729,~\omega_2=3.93084,~\omega_3=4.05,~k=0.1,~F=1.08,~G=0.7$. The entrainment window in $\alpha$ is [0.285:0.497].}
	\label{figew3}
\end{figure}

The fixed points are $A_1^*,~A_2^*,~B_1^*,~B_2^*$ with $r_1^2=(A_1^*)^2+(B_1^*)^2$,$r_2^2=(A_2^*)^2+(B_2^*)^2$, given by the following relations
\begin{subequations}
  \begin{align}
 &(\omega_1^2-\omega_3^2)B_1^* - \alpha B_2^* = k\omega_3(\frac{r_1^2}{4}-1)A_1^* \label{46a}\\
 &(\omega_1^2-\omega_3^2)A_1^* -\alpha A_2^* - k\omega_3(\frac{r_1^2}{4}-1)B_1^* = F \\
 &(\omega_2^2-\omega_3^2)B_2^* - \alpha B_1^* = k\omega_3(\frac{r_2^2}{4}-1)A_2^* \\
 &(\omega_2^2-\omega_3^2)A_2^* - \alpha A_1^* + k\omega_3(\frac{r_2^2}{4}-1)B_2^* = G. \label{46d} 
  \end{align}
\end{subequations}
Finally, the stability of the fixed points can be found from
\begin{subequations}
  \begin{align}
 &2\omega_3\delta \dot{A}_1=\left[\omega_1^2-\omega_3^2-A_1^*B_1^*\frac{k\omega_3}{2}\right]\delta B_1-\alpha \delta B_2 \label{47a}\nonumber\\
 &+ k\omega_3\left[1-\left(\frac{3}{4}(A_1^*)^2 +\frac{1}{4}(B_1^*)^2\right)\right]\delta A_1\\
 & 2\omega_3\delta \dot{B}_1=-\left[\omega_1^2-\omega_3^2+A_1^*B_1^*\frac{k\omega_3}{2}\right]\delta A_1+\alpha \delta A_2\nonumber\\
 & +k\omega_3\left[1-\left(\frac{3}{4}(B_1^*)^2 +\frac{1}{4}(A_1^*)^2\right)\right]\delta B_1\\
 &2\omega_3\delta \dot{A}_2=\left[\omega_2^2-\omega_3^2-A_2^*B_2^*\frac{k\omega_3}{2}\right]\delta B_2-\alpha \delta B_1\nonumber\\
 &+ k\omega_3\left[1-\left(\frac{3}{4}(A_2^*)^2 +\frac{1}{4}(B_2^*)^2\right)\right]\delta A_2\\ 
  & 2\omega_3\delta \dot{B}_2=-\left[\omega_2^2-\omega_3^2+A_2^*B_2^*\frac{k\omega_3}{2}\right]\delta A_2+\alpha \delta A_1\nonumber\\
 & +k\omega_3\left[1-\left(\frac{3}{4}(B_2^*)^2 +\frac{1}{4}(A_2^*)^2\right)\right]\delta B_2. \label{47d}
  \end{align}
\end{subequations}
The systems decouple for $\alpha=0$ and the entrainment regions for a single forced Van der Pol Oscillator are shown in Jordan and Smith \cite{JS}. Hence, we ask the question that if for $\alpha=0$, one of the oscillator is in the entrained state and the other is not then is it possible for them to be synchronised for $\alpha\neq 0$. In principle, an approximate answer to the question can be found by obtaining the fixed points from equations \eqref{46a}-\eqref{46d} and testing their stability using equations \eqref{47a}-\eqref{47d}. However, in this work, we opt for the direct process of solving equations \eqref{qE}-\eqref{qE_b} numerically to obtain the stability zone if any.\\
\indent
We choose the forcing frequency as $\omega_3=4$ and the amplitude $F=1.08$, $G=0.78$ when $\alpha=0$. For this frequency the forced Van der Pol oscillator $\ddot{x}+k\dot{x}(x^2-1)+\omega_1^2x=Fcos(\omega_3 t)$ is entrained. We choose $\omega_2$ in such a way that the second oscillator lies outside the entrainment zone. Now we switch on the coupling $\alpha$. Since one oscillator is entrained to the external forcing and the other is not, the two oscillators are not synchronized for small $\alpha$. However, at a critical value $\alpha_1$, we find both the oscillators to be entrained to the external forcing. This entrainment stays for a range of $\alpha$ but for $\alpha$ greater than a critical value, $\alpha_2$, the oscillators are out of entrainment once again. The loss of entrainment on increasing $\alpha$ beyond $\alpha_2$ is possibly a reflection of the separation of the natural frequencies of the normal modes of the coupled systems (in the absence of nonlinear term, i.e. for $k=0$). The variation of the entrainment zone is quite sensitive to changes in the frequency and amplitude of the external forcing. If we keep $F=1.08$ and $G=0.7$ as in Fig. \ref{figew} and change the frequency of the drive to $\omega_3=4.05$, the entrainment window (see Fig. \ref{figew3}) is significantly changed to $0.285<\alpha<0.497$.  
The upper limit of the coupling constant, $\alpha$ is $=\omega_1\omega_2$ ($\because~\Omega_2$ has to be real). 
\subsection{Dependence of the Entrainment Window  on the Parameters}
\subsubsection{In the absence of coupling, one of the oscillator is entrained at the frequency of the external forcing but the other is not} 
We keep the parameters of the first oscillator ($\omega_1,~k,~F$)  fixed such that in the absence of coupling, it is entrained at the frequency of the external periodic forcing. Obviously, we also keep the frequency of the external periodic forcing, $\omega_3$  fixed. We find the behaviour of the entrainment window (the width and the location) with the variation of the parameters of the second oscillator which is always not entrained at the frequency of the external forcing within that variation. Therefore, to do this,  just as before, we vary the coupling $\alpha$  between them while setting the amplitude of the external periodic forcing on the second oscillator, $G$ and $\omega_2$ as two parameters of the problem. In the absence of coupling, the parameters of the second oscillator are $n_2(=\frac{\omega_3^2-\omega_2^2}{k\omega_3}),~g_2(=\frac{G}{k\omega_3})$ . 
We vary coupling to see how coupling between them can lift the second oscillator to the stable region while the first oscillator is entrained at the frequency of the external periodic forcing (the location of the first oscillator is fixed the stable region of the stablity diagram \cite{JS}). We seek to find out the entrainment window with the variation of these two parameters.
\begin{figure}[hbtp]
	\centering
    \includegraphics[scale=0.35]{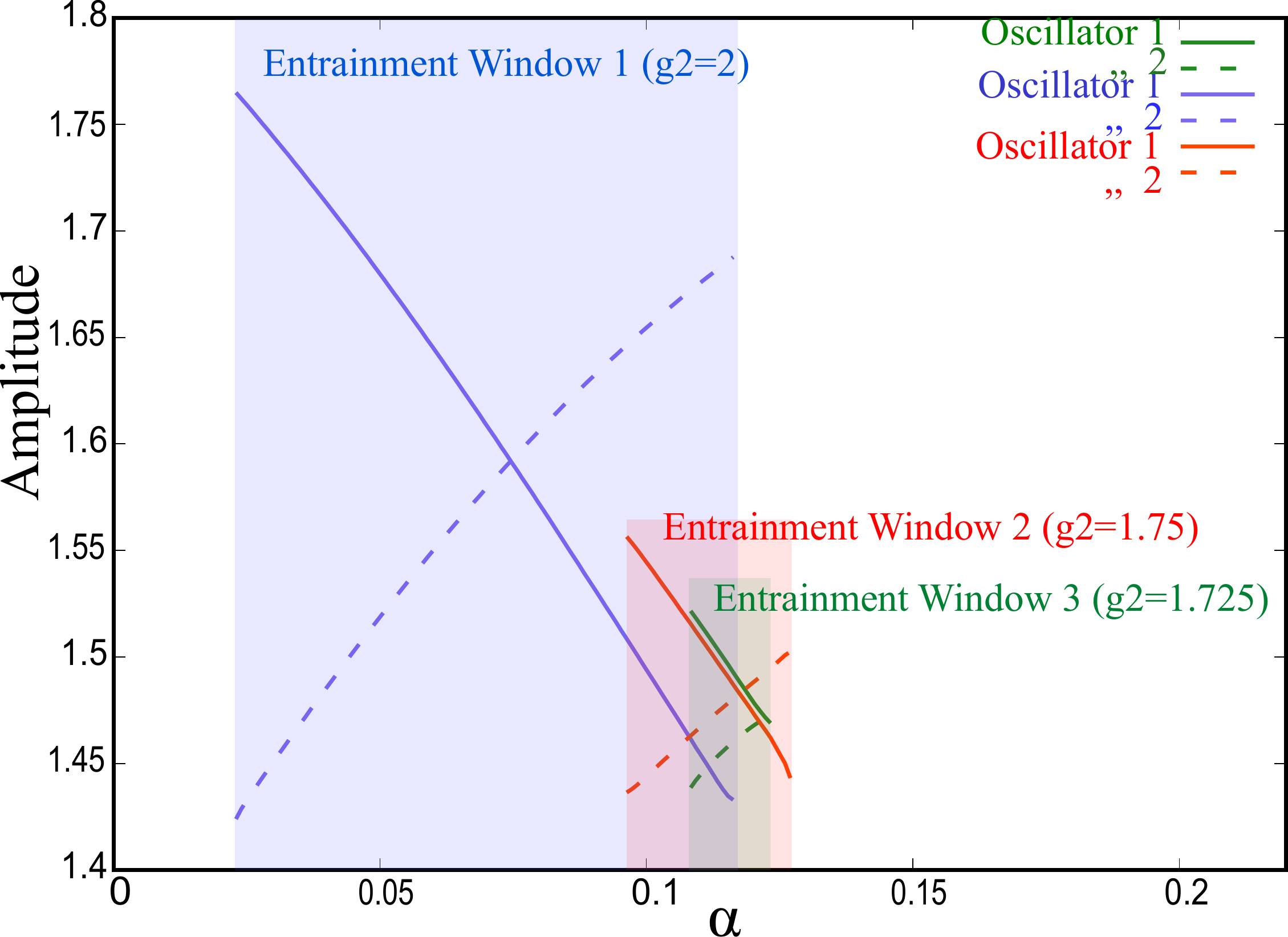}
	\caption{With the  increase in forcing amplitude on the second oscillator, the entrainment window widens  with a tendency to shift its position to the left on the $\alpha$ axis. The entrainment window diminishes after a certain lower limit in $g_2~(=1.725)$ within our numerical precision. The fixed coefficients: $\omega_3=4$, $\omega_1=4.0729$, $k=0.1$, $F=1.08$, $\omega_2=3.93$.}
	\label{figew4}
\end{figure}
\begin{figure}[hbtp]
	\centering
    \includegraphics[scale=0.35]{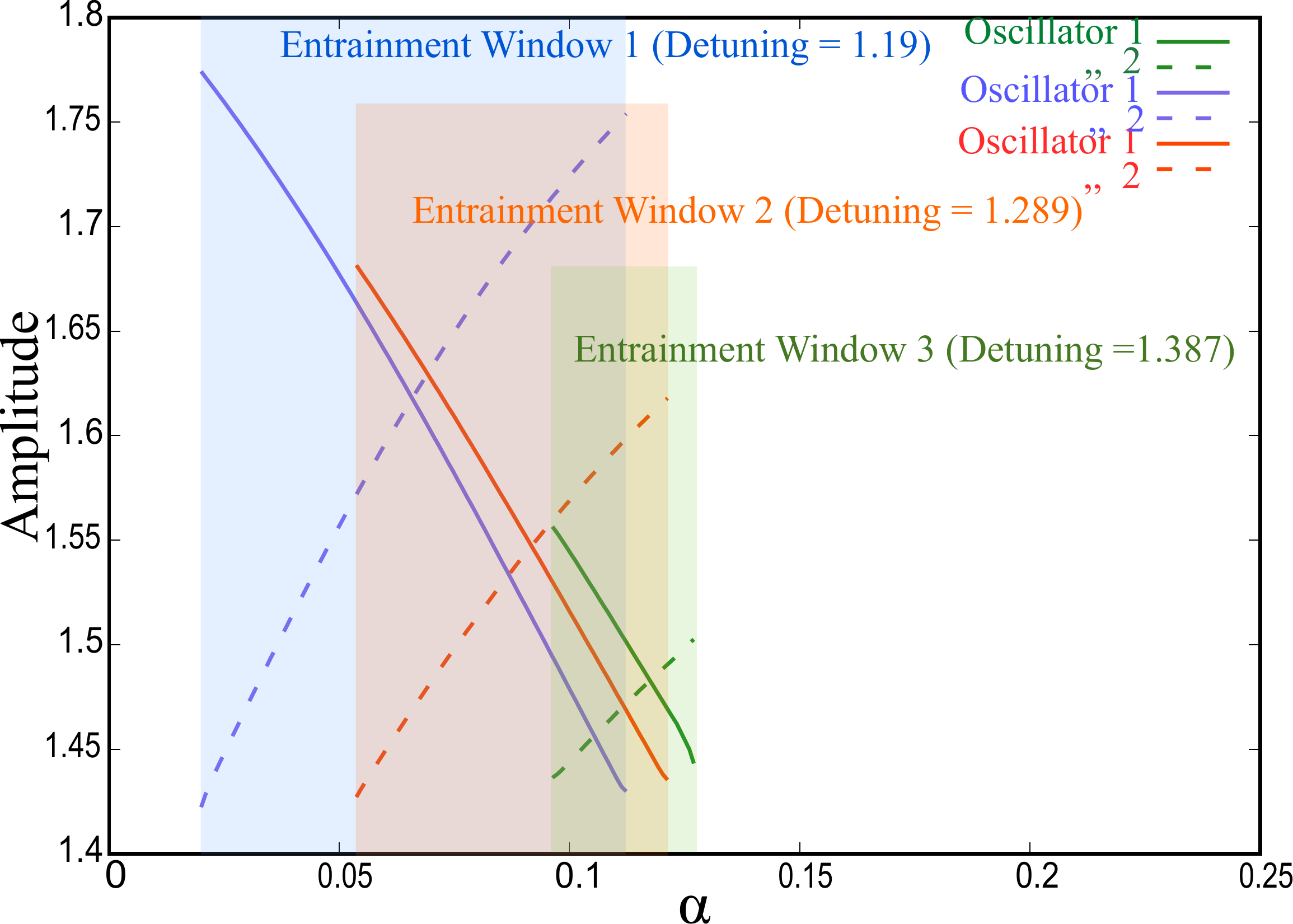}
	\caption{With the  decrease in detuning of the second oscillator, the entrainment window widens  with a tendency to shift its position to the left on the $\alpha$ axis. The entrainment window diminishes after a certain upper limit in detuning, $n_2~(=1.387)$ within our numerical precision. The fixed coefficients: $\omega_3=4$, $\omega_1=4.0729$, $k=0.1$, $F=1.08$, $G=0.7$. }
	\label{figew5}
\end{figure}
The events of figure \ref{figew4} and figure \ref{figew5} can be qualitatively explained as follows. 
 With the increase in $g_2$, the location of the second oscillator in the  unstable region of the stability diagram (in the uncoupled state), shifts towards the boundary between stable and unstable region, hence making it easier (with smaller value of coupling) for the first oscillator (the entrained one in the absence of coupling) to make the coupled system entrained, and it turns out that the entrainment last for larger range in $\alpha$. Thus the entrainment window shifts to the left and also widens with the increase in $g_2$. 
 Similar qualitative explanation holds for the figure \ref{figew5}.   
\subsubsection{In the absence of coupling, both of the oscillators are entrained at the frequency of the external forcing} 
\begin{figure}[hbtp]
 	\centering
   \includegraphics[scale=0.35]{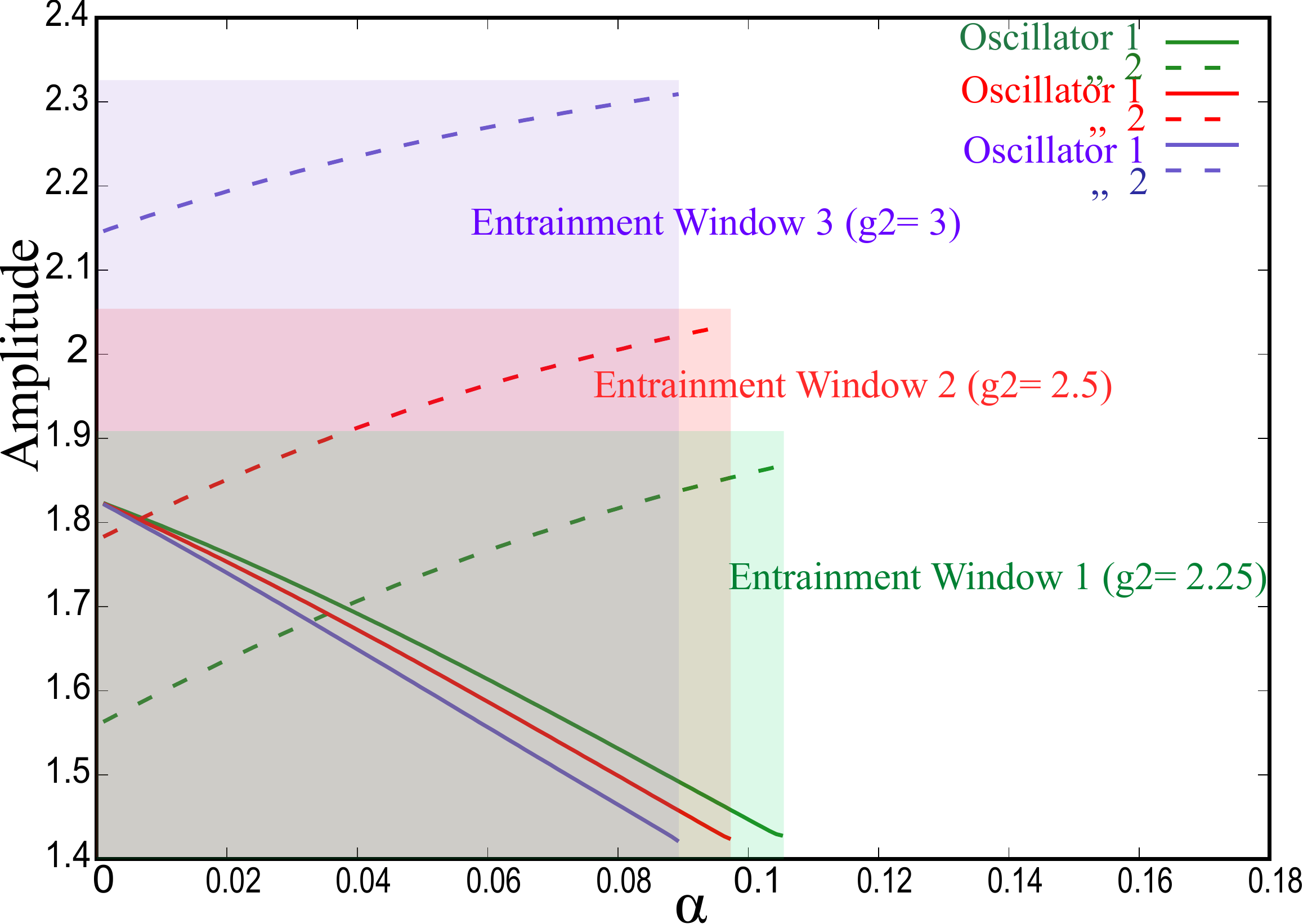}
 	\caption{With the  increase in forcing amplitude on the second oscillator, the entrainment shifts its position to the left on the $\alpha$ axis. The fixed coefficients: $\omega_3=4$, $\omega_1=4.0729$, $k=0.1$, $F=1.08$, $\omega_2=3.93$. }
 	\label{figew6}
 \end{figure}

\begin{figure}[hbtp]
 	\centering
     \includegraphics[scale=0.35]{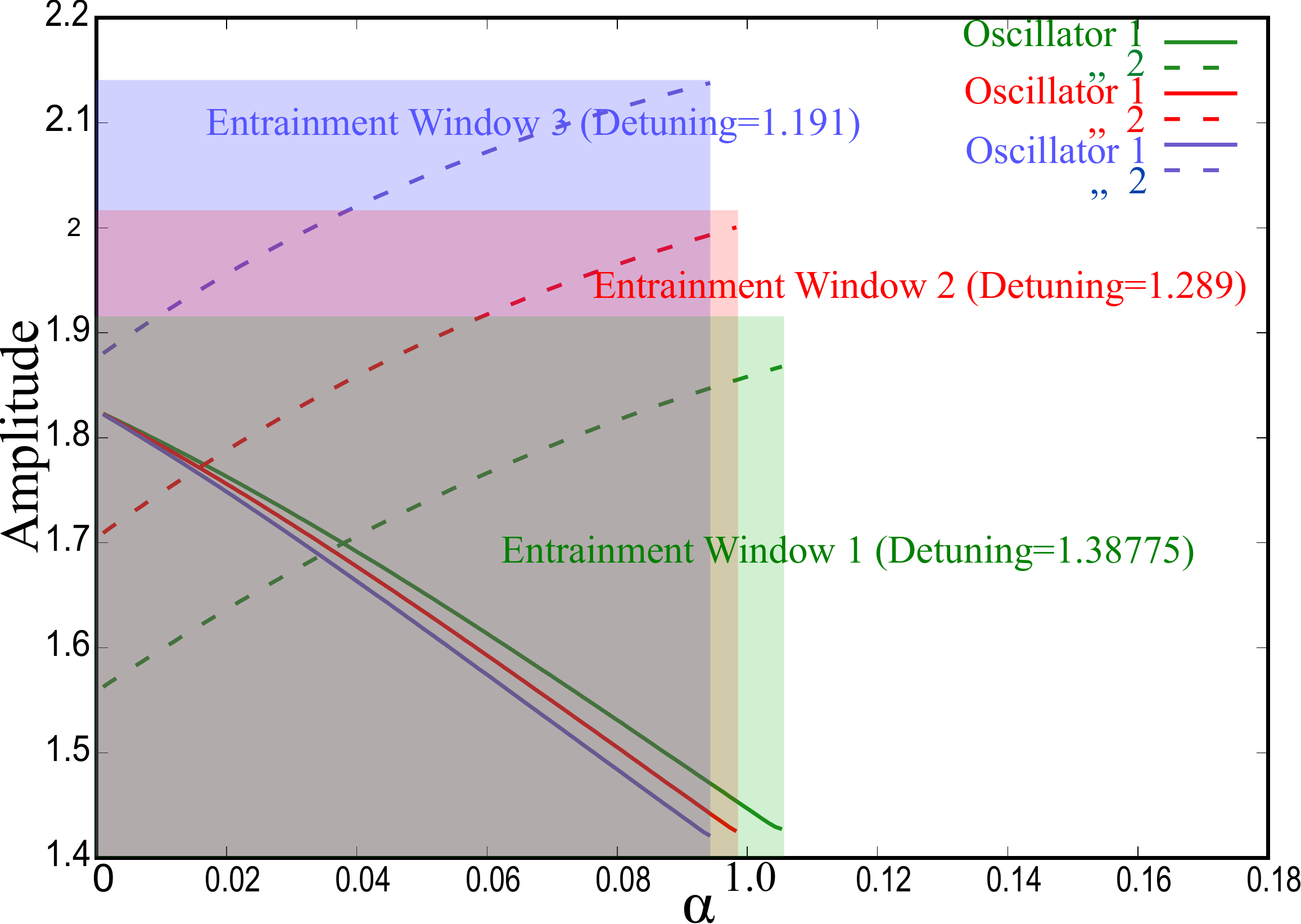}
 	\caption{With the  decrease in detuning of the second oscillator, the entrainment window shifts its position to the left on the $\alpha$ axis. The fixed coefficients: $\omega_3=4$, $\omega_1=4.0729$, $k=0.1$, $F=1.08$, $G=0.9$. }
 	\label{figew7}
 \end{figure}

Let's consider the case where both the oscillators are individually entrained in the absence of coupling (for $\alpha=0$). Interestingly, we see in the figure \ref{figew6} and figure \ref{figew7} that after certain value of $\alpha$, the system goes out of sync with the external periodic forcing. 
Thus coupling destroys entrainment of the system. The entrainment window in $\alpha$ shifts towards left with the increase in the forcing amplitude and with the decrease in detuning of the second oscillator. 
In the absence of coupling, if both of the oscillators are out of sync with the frequency of the external forcing, coupling can not create entrainment in such system. We have checked this fact for a set of parameters. Although we have not found any analytical reasoning.
\section{Conclusion} 
Limit cycles are generally characterized as isolated closed curves. In higher dimensional systems we
show that the closed curve can be generalized to isolated closed figures like a torus. We have
explored systems characterized by two incommensurate frequencies- a pair of coupled Van der Pol
oscillators and the Kuznetsov oscillator. It is seen that in such settings the renormalization group
formulation of perturbation theory is the most natural tool to use. This has been demonstrated in
Secs II and III. In particular our treatment of the Kuznetsov oscillator seems to be the first attempt at
analytically characterizing this limit cycle. In Sec IV we return to the coupled Van der Pol oscillators
but now under external forcing. Our most significant finding is that if one of the oscillators is
synchronized to the external forcing and the other is not , it is still possible to induce synchronization
in the system by means of the mutual coupling.
    
\section{Acknowledgement}
DKM acknowledges financial support by the ANR project Dirac3D under grant no. ANR-17-CE30-0023 ``Dirac 3D". The research of SD was supported in part by the INFOSYS scholarship for senior students. SD is thankful to Dr. Jayita Lahiri for useful discussions and for helping him in a numerical coding.
\section{Appendix}

In this appendix, we exhibit the intermediate steps leading to the renormalization of frequency in Eqs. (\ref{eq:RGdefinition_b}) and (\ref{eq:RGdefinition_d}). This requires carrying on the perturbation theory developed in Sec.~\ref{sec:cvpo} to the next order. At the end of the first order of perturbation theory, we had arrived at
\begin{subequations}
  \begin{align}
    x_{00} &= A\cos(\omega_1t+\theta) \\
    x_{10} &= -\frac{A^3}{32\omega_1}\sin 3(\omega_1t+\theta)\\
    x_{01} &= \frac{B}{\omega_1^2-\omega_2^2}\cos (\omega_2t+\theta')\\
    y_{00} &= B\cos(\omega_2t+\theta') \\
    y_{10} &= -\frac{B^3}{32\omega_2}\sin 3(\omega_2t+\theta')\\
    y_{01} &= \frac{A}{\omega_2^2-\omega_1^2}\cos (\omega_1t+\theta)
  \end{align}
\end{subequations}

To arrive at the equation of motion for $x_{20}$, we need the following quantities
\begin{subequations}
  \begin{align}
    \dot{x}_{10}&=-\frac{3}{32}A^3\cos 3(\omega_1t+\theta) \\
    x_{00}^2x_{10}&=\frac{A^5}{128\omega_1}\{\sin(\omega_1t+\theta) \nonumber \\ & \hspace{1cm} + 2\sin 3(\omega_1t+\theta) + \sin5(\omega_1t+\theta)\}
  \end{align}
\end{subequations}
with
\begin{align}
  \frac{\rm{d}}{\rm{dt}}\{x_{00}^2x_{10}\} =& \frac{A^5}{128}\{\cos(\omega_1t+\theta) + 6\cos 3(\omega_1t+\theta) \nonumber \\ & \hspace{1cm} + 5\cos 5(\omega_1t+\theta)\}.
\end{align}

This leads to
\begin{align}
  \ddot{x}_{20}+\omega_1^2x_{20}=&-\frac{3}{32}A^3\cos 3(\omega_1t+\theta) + \frac{A^5}{128}\{\cos(\omega_1t+\theta)\nonumber \\ & \hspace{0cm} + 6\cos 3(\omega_1t+\theta) + 5\cos 5(\omega_1t+\theta)\}
\end{align}
with the particular integral
\begin{align}
  x_{20}=&\frac{A^5}{256\omega_1}(t-t_0)\sin\omega_1(t-t_0) +\frac{3}{512}A^3\cos 3(\omega_1t+\theta)\nonumber \\ & \hspace{0cm}-\frac{5 A^5}{24\times 128}\cos 5(\omega_1t+\theta).
\end{align}

Only the first terms in $x_{20}$ above will diverge at long times. Proceeding further,
\begin{align}
  \dot{x}_{01}(x_{00}^2-1) &= \frac{B}{\omega_2^2-\omega_1^2}\omega_2\sin(\omega_2t+\theta') \nonumber \\ & \hspace{0cm} \times\{\frac{A^2}{2}-1+\frac{A^2}{2}\cos 2(\omega_1t+\theta)\} \\
  x_{01}\frac{\rm{d}}{\rm{dt}}x_{00}^2 &= -\frac{A^2B}{\omega_1^2-\omega_2^2}\omega_1\cos (\omega_2t+\theta')\sin 2(\omega_1t+\theta)
\end{align}
and
\begin{align}
  y_{10} = -\frac{B^3}{32\omega_2}\cos 3(\omega_2t+\theta').
\end{align}

Evidently there are no resonating terms in the dynamics of $x_{11}$ or $y_{11}$. Showing (?) the resonating term in the dynamics of $x_{20}$, we have
\begin{align}
  \ddot{x}_{02}+\omega_1^2x_{02}= \frac{A}{\omega_2^2-\omega_1^2}\cos(\omega_1t+\theta)
\end{align}
with the particular integral
\begin{align}
  x_{02} = \frac{A}{2\omega_1}\frac{1}{\omega_2^2-\omega_1^2}(t-t_0)\sin\omega_1(t-t_0).
\end{align}

We now write down the solution for $x(t)$ at this stage showing the complete solution upto the first order and only the divergent terms at the second order. This gives
\begin{align}
  x(t) = &A\cos(\omega_1t+\theta) \nonumber \\ & \hspace{0cm} + \frac{\alpha B}{\omega_1^2-\omega_2^2}\cos(\omega_2t+\theta')-\frac{kA^3}{32\omega_1}\cos 3(\omega_1t+\theta) \nonumber\\ &+ \frac{\alpha^2 A}{2\omega_1}\frac{1}{\omega_2^2-\omega_1^2}(t-t_0)\sin\omega_1(t-t_0) \nonumber \\ & \hspace{0cm} - \frac{k^2A^5}{256\omega_1}(t-t_0)\sin\omega_1(t-t_0) + \rm{non~divergent~terms}.
\end{align}

The renormalization necessary to remove the divergence in Eq.~(?) has to be in the phase and accordingly we write the phase as $\theta+\alpha^2\tilde{Z}_{02}$, so that $\cos(\omega_1t+\theta)$ becomes $\cos(\omega_1t+\theta+\alpha^2\tilde{Z}_{02}+k^2\tilde{Z}_{20})$ given by
\begin{align}
 & \cos(\omega_1t+\theta+\alpha^2\tilde{Z}_{02}+k^2\tilde{Z}_{20})\nonumber \\ & \hspace{0cm} = \cos(\omega_1t+\theta) - \alpha^2\tilde{Z}_{02}\sin(\omega_1t+\theta)- k^2\tilde{Z}_{20}\sin(\omega_1t+\theta) \nonumber \\ & \hspace{1cm} + \mathcal{O}(\alpha^4).
\end{align}

Showing only the $\mathcal{O}(\alpha^2)$ and $\mathcal{O}(k^2)$ corrections in $x(t)$, we have
\begin{align}
  x(t) =& A\cos(\omega_1t+\theta) - A\alpha^2\tilde{Z}_{02}\sin(\omega_1t+\theta) \nonumber \\ & \hspace{0cm} - A k^2\tilde{Z}_{20}\sin(\omega_1t+\theta) \nonumber \\ & \hspace{0cm} + \frac{\alpha^2 A}{2\omega_1(\omega_2^2-\omega_1^2)}(t-\tau+\tau-t_0)\sin(\omega_1t+\theta) \nonumber \\ & \hspace{0cm}- \frac{k^2 A^5}{256\omega_1}(t-\tau+\tau-t_0)\sin(\omega_1t+\theta)+\rm{finite~terms}.
\end{align}

Removal of `$t_0$' requires
\begin{subequations}
  \begin{align}
    \tilde{Z}_{02} &= \frac{1}{2\omega_1}\frac{1}{\omega_2^2-\omega_1^2}(\tau - t_0) \\
    \tilde{Z}_{20} &= -\frac{A^4}{256\omega_1}(\tau - t_0).
  \end{align}
\end{subequations}

This leaves
\begin{align}
  x = & A\cos(\omega_1t+\theta)+\frac{\alpha B}{\omega_1^2-\omega_2^2}\cos(\omega_2t+\theta')\nonumber \\& -\frac{kA^3}{32\omega_1}\cos 3(\omega_1t+\theta) - \frac{k^2A^5}{256\omega_1}(t-\tau)\sin(\omega_1t+\theta)\nonumber \\&  +\frac{\alpha^2A}{2\omega_1}\frac{1}{\omega_2^2-\omega_1^2}(t-\tau)\sin(\omega_1t+\theta).
\end{align}

The requirement that $x(t,\tau)$ is independent of $\tau$ leads to
\begin{align}
  \frac{\rm{d}\theta}{\rm{d}\tau} = -\frac{\alpha^2}{2\omega_1}\frac{1}{\omega_2^2-\omega_1^2} + \frac{k^2A^4}{256\omega_1}.\label{eq:RGequations_Appendix_theta}
\end{align}

At this second order of calculation, there is no amplitude renormalization and hence choose $\tau = t$, in Eq.~(\ref{eq:RGequations_Appendix_theta})
\begin{subequations}
  \begin{align}
    \frac{\rm{d}\theta}{\rm{d}t} &= -\frac{\alpha^2}{2\omega_1}\frac{1}{\omega_2^2-\omega_1^2} + \frac{k^2A^4}{256\omega_1}\\
    \frac{\rm{d}A}{\rm{d}t} &= \frac{kA}{2}(1-\frac{A^2}{4})+\mathcal{O}(k^3).
  \end{align}
\end{subequations}

This shows that at $\mathcal{O}(\alpha^2,k^2)$
\begin{align}
  A^2 &= 4 + \mathcal{O}(k^2)\\
  \theta &= -\frac{\alpha^2}{2\omega_1}\frac{1}{\omega_2^2-\omega_1^2}t + \frac{k^2t}{16\omega_1}.
\end{align}

The frequency $\omega_1$ is consequently renormalized to
\begin{align}
  \omega_{1R} = \omega_1 - \frac{\alpha^2}{2\omega_1}\frac{1}{\omega_2^2-\omega_1^2} + \frac{k^2}{16\omega_1} + \rm{higher~order~terms}
\end{align}
and similarly, the frequency $\omega_2$ renormalizes to 
\begin{align}
  \omega_{2R} = \omega_2 - \frac{\alpha^2}{2\omega_2}\frac{1}{\omega_1^2-\omega_2^2} + \frac{k^2}{16\omega_2} + \rm{higher~order~terms}.
\end{align}

\newpage  
\bibliographystyle{ieeetr}
\bibliography{References}

\end{document}